\documentclass[preprint,12pt]{elsarticle}

\usepackage{graphicx, epsfig, subfigure}
\graphicspath{{figures/}}

\begin{document}

\begin{frontmatter}

\title{Systematic sensitivity study of the $J/\psi$ nuclear modification factor to polarization assumptions}

\author[1,2]{Yi Yang\corref{cor1}}
\ead{yiyang429@as.edu.tw}
\author[2]{Chun-Wei Su}
\author[1,2]{Te-Chuan Huang}

\address[1]{Institute of Physics, Academia Sinica, Taiwan, ROC}
\address[2]{Department of Physics, National Cheng Kung University, Taiwan, ROC}

\cortext[cor1]{Corresponding author}

\begin{abstract}
Heavy quarkonium is one of the key candidates to study the fundamental properties of Quark-Gluon Plasma (QGP) created in heavy-ion ($A$+$A$) collisions. Comparing the production of the $J/\psi$ meson in proton+proton ($p$+$p$) and $A$+$A$ collisions, namely the nuclear modification factor ($R_{\rm AA}$), provides the quantitative understanding of the QGP. Normally, the $R_{\rm AA}$ is measured under the assumption that the quarkonium is unpolarized. However, recent measurements on the $J/\psi$ polarization in the forward rapidity region from the LHC experiments suggest that the $J/\psi$ meson has a small but non-negligible polarization. In this paper, we evaluate the systematic sensitivity of the kinematic acceptance to the $J/\psi$ polarization assumptions in the measurement of $R_{\rm AA}$. Using available data from the ALICE and LHCb experiments in the forward rapidity region, and exploring extreme polarization scenarios in the central rapidity region at RHIC and LHC energies, we demonstrate the maximum envelope of systematic uncertainties induced by unknown polarization states. The results indicate that the unpolarized assumption introduces a significant systematic uncertainty. Having precise measurements of quarkonium polarization in heavy-ion collisions is critical to obtaining an accurate interpretation of how heavy quarkonium interacts with the QGP.
\end{abstract}

\begin{keyword}
$J/\psi$ meson \sep nuclear modification factor \sep Quark-Gluon Plasma \sep polarization \sep kinematic acceptance
\end{keyword}

\end{frontmatter}

\section{Introduction}
Heavy quarkonia, the heavy flavor quark-antiquark pairs, are important candidates to understand the fundamental properties of Quantum Chromodynamics (QCD). Studying the production of heavy quarkonium in proton-proton ($p$+$p$) and heavy-ion ($A$+$A$) collisions provides fruitful information on QCD since it covers both the perturbative (hard scattering) and non-perturbative (hadronization stage) regions, as well as the knowledge of the new state of matter, Quark-Gluon Plasma (QGP), which is expected to be created in $A$+$A$ collisions.

The $J/\psi$ meson is the $c\bar{c}$ bound state which was discovered in 1974~\cite{jpsi_1, jpsi_2} and it often serves as a standard candle for the studies of quarkonium properties. Measuring the $J/\psi$ nuclear modification factor, $R_{\rm AA}$, which is the ratio of the invariant production yields of $J/\psi$ in $A$+$A$ collisions normalized to the number of binary collisions to that in $p$+$p$ collisions, is one of the important ways to provide deep understanding of the QGP. The $R_{\rm AA}$ for the $J/\psi$ meson in $A$+$A$ collisions is defined as 
\begin{eqnarray}
    \label{eq:RAA}
    R_{\rm AA} = \frac{1}{\langle N_{coll} \rangle}\frac{\left( \frac{1}{2\pi p_T}\frac{d^2N_{J/\psi}}{dy dp_T}\right)_{A+A}}{\left( \frac{1}{2\pi p_T} \frac{d^2N_{J/\psi}}{dy dp_T}\right)_{p+p}},    
\end{eqnarray}
where $\langle N_{coll} \rangle$ is the average number of binary nucleon-nucleon collisions and $\left( \frac{1}{2\pi p_T}\frac{d^2N_{J/\psi}}{dy dp_T}\right)_{A+A\ (p+p)}$ is the invariant yield of the $J/\psi$ meson in $A$+$A$ ($p$+$p$) collisions. The invariant yield can be expressed as
\begin{eqnarray}
    \frac{d^2 N_{J/\psi}}{2\pi p_T dp_T dy} = \frac{N^{\rm raw}_{J/\psi}}{(2\pi p_T)\cdot \mathcal{A} \cdot \varepsilon  \cdot \Delta p_T \cdot \Delta y}, 
\label{eq:inv_yield}
\end{eqnarray}
where $N^{\rm raw}_{J/\psi}$ is the raw number of reconstructed $J/\psi$; $\mathcal{A}$ is the $J/\psi$ kinematic acceptance which is defined as the ratio of the number of events passed certain kinematic criteria based on the detector configuration to the number of events without any restrictions; $\varepsilon$ is the reconstruction efficiency; and $\Delta p_T$ and $\Delta y$ are the corresponding bin widths. 

It is strictly necessary to note that the angular distributions of the decayed leptons from $J/\psi$ are heavily dependent on the polarization of $J/\psi$ as described in the following:
\begin{eqnarray}
    \label{eq:polar}
    W(\cos\theta, \phi) \propto  (1 + \lambda_{\theta}\cos^2\theta + \lambda_{\phi}\sin^2\theta\cos2\phi + \lambda_{\theta\phi}\sin2\theta \cos\phi),     
\end{eqnarray}
where $\theta$ is the polar angle, $\phi$ is the azimuthal angle, and $\lambda_{\theta}$, $\lambda_{\phi}$, and $\lambda_{\theta\phi}$ are the polarization parameters (detailed definitions can be found in Ref.~\cite{quark_pol_def}). Consequently, the kinematic acceptance $\mathcal{A}$ of $J/\psi$ in Eq.~\ref{eq:inv_yield} is an explicit function of the $J/\psi$ polarization.

Furthermore, the polarization states of prompt $J/\psi$ (direct production or feed-down from higher excited states) and non-prompt $J/\psi$ (from $B$-hadron weak decays) are fundamentally different. Applying a universal unpolarized assumption oversimplifies the physical complexity of the measured inclusive spectra.

Obtaining a rigorous interpretation of the $R_{\rm AA}$ measurement requires extra care since it is influenced by both hot nuclear matter effects (HNM)~\cite{hnm_1, hnm_2} and cold nuclear matter effects (CNM)~\cite{cnm}, such as nuclear shadowing and the Cronin effect. Normally, the kinematic acceptance correction for invariant yields is blindly based on an unpolarized $J/\psi$ assumption. However, recent ALICE measurements on the $J/\psi$ polarization in the forward rapidity ($y$) region in Pb+Pb collisions at $\sqrt{s_{\rm NN}} = $ 5.02 TeV suggest that the $J/\psi$ meson is slightly transversely and longitudinally polarized in the helicity (HX) and Collins-Soper (CS) frames at low $p_T$, respectively~\cite{alice_jpsi_pol_pbpb}; while measurements from the LHCb experiment show that $J/\psi$ is slightly longitudinally polarized in both frames in $p$+$p$ collisions at $\sqrt{s} =$ 7 TeV~\cite{lhcb_jpsi_pol_pp}. 

These experimental facts indicate that heavy quarkonium has distinct production mechanisms or varying methods of acquiring polarization in different collision systems. Therefore, implementing kinematic acceptance with the appropriate polarization boundary is crucial. In this paper, we conduct a systematic sensitivity study of the polarization effect on the kinematic acceptance to establish the systematic boundary limits of the $J/\psi$ $R_{\rm AA}$ measurements using available data from LHC and RHIC.

\section{Analysis Procedure} 
This analysis evaluates the systematic sensitivity using Toy Monte Carlo (MC) samples implemented with the exact kinematic configurations from the LHC and RHIC experiments. The quantitative study of the effect from the $J/\psi$ kinematic acceptance under varied polarization parameters is achieved by comparing the input $p_T$ distribution in Toy MC events to the $p_T$ distribution from events corrected using the unpolarized kinematic acceptance.

Firstly, the polarization parameters in both HX and CS frames for the $J/\psi$ meson ($\lambda_{\theta}$, $\lambda_{\phi}$, and $\lambda_{\theta\phi}$) are extracted from measurements. Secondly, high-statistics $J/\psi$ events ($10^8$) are generated using a single-particle generator, forcing $J/\psi$ to decay into a dilepton pair. The input $p_T$ and $y$ spectra of $J/\psi$ are strictly obtained from real measurements and global constraints~\cite{jpsi_global}. The polarization of the $J/\psi$ meson is assigned via $p_T$-dependent parameters parameterized from measurements, following the angular distribution in Eq.~\ref{eq:polar}.

Next, identical kinematic selections from the real measurements are applied. A candidate-by-candidate weighting method is used to correct the kinematic acceptance, defined as $N^{\rm corr.} = \sum_{i=1}^{N_{J/\psi}} w_i$, where $w_i = 1/\mathcal{A}$~\cite{atlas_jpsi, atlas_upsi, star_jpsi}.

It is important to emphasize a known experimental limitation of this approach. This study isolates the effect of the pure kinematic acceptance ($\mathcal{A}$). In real experimental conditions, the detector reconstruction efficiency ($\varepsilon$) strongly depends on the specific kinematics of the single decay muons, particularly at low $p_T$, which inevitably couples with the polarization state. Since simulating the full detector response is beyond the scope of this phenomenological study, the variations presented here reflect the first-order approximation purely from kinematic acceptance boundaries. Consequently, the derived systematic uncertainty acts as a conservative lower bound, as full detector efficiency coupling could potentially amplify the discrepancies.

The effect on the $J/\psi$ invariant yield originated from the kinematic acceptance utilizing inaccurate unpolarized assumptions is quantified by the ratio of the input $p_T$ spectrum to the acceptance-corrected one. The sensitivity boundary of the modified $R_{\rm AA}$, denoted as $R_{\rm AA}^{\rm corr.}$, can be expressed as:
\begin{eqnarray}
    R_{\rm AA}^{\rm corr.} &=&  R_{\rm AA} \times \frac{C_{\rm AA}^{p_T, y}}{ C_{\rm pp}^{p_T, y}} \nonumber \\
&=& \frac{1}{\langle N_{coll} \rangle}\frac{\left( \frac{1}{2\pi p_T}\frac{d^2N_{J/\psi}}{dp_T dy}\right)_{A+A}  }{\left( \frac{1}{2\pi p_T} \frac{d^2N_{J/\psi}}{dp_T dy}\right)_{p+p}} \times \frac{ \left( \frac{N^{\rm input}}{N^{\rm corr}.} \right)_{A+A}^{p_T, y} }{  \left( \frac{N^{\rm input}}{N^{\rm corr.}} \right)_{p+p}^{p_T, y}  },
    \label{eq:raa_corr}
\end{eqnarray}
where $C_{\rm AA (pp)}^{p_T, y}$ is the corresponding correction factor in $A$+$A$ ($p$+$p$) collisions. It is defined as the ratio of $N^{\rm input}$ (events without kinematic selections) to $N^{\rm corr.}$ (events with kinematic selections corrected by unpolarized assumptions). The denominator in $C_{\rm AA (pp)}^{p_T, y}$ cancels out the unpolarized kinematic acceptance used in published measurements.

\section{Results}

Before presenting the results in the different kinematic regions, we clarify the interpretation of the polarization-frame dependence in this study. If an identical and fully specified physical polarization state is transformed exactly between polarization frames, the corresponding kinematic acceptance correction should be frame independent. The present work, however, does not perform such an exact frame transformation of a single underlying polarization state. Instead, it uses the available experimental polarization measurements as phenomenological inputs in each frame.

In the forward-rapidity region, the input parameters $\lambda_\theta$, $\lambda_\phi$, and $\lambda_{\theta\phi}$ are obtained from independent parameterizations of the published measurements in the HX and CS frames. Owing to the experimental uncertainties and the limited one-dimensional $p_T$-dependent parameterization used here, these two sets of inputs do not necessarily correspond to the exact same physical polarization state expressed in two different frames. The resulting HX and CS calculations should therefore be interpreted as frame-dependent sensitivity estimates constrained by the available measurements, rather than as exact transformations of one unique polarization density matrix. This effect is particularly visible at forward rapidity, where the non-uniform and asymmetric muon acceptance enhances the sensitivity to the decay angular distribution.

For the central-rapidity calculations, the extreme polarization scenarios are also imposed independently in the HX and CS frames in order to estimate the maximum possible systematic envelope in the absence of direct heavy-ion polarization measurements. Although these imposed scenarios are not exact frame transformations of one another, the more symmetric central-rapidity geometrical acceptance reduces the sensitivity of the integrated correction factors to the frame choice. This explains why the HX and CS correction factors are numerically similar in Figs.~8--11, while larger differences can appear in the forward-rapidity results shown in Figs.~4 and 5.

\subsection{Forward rapidity region}
\begin{figure}[!htbp]
  \begin{center}
      \subfigure[]{
      \label{fig:hx_polar_lhc}
      \includegraphics[width=0.47\textwidth]{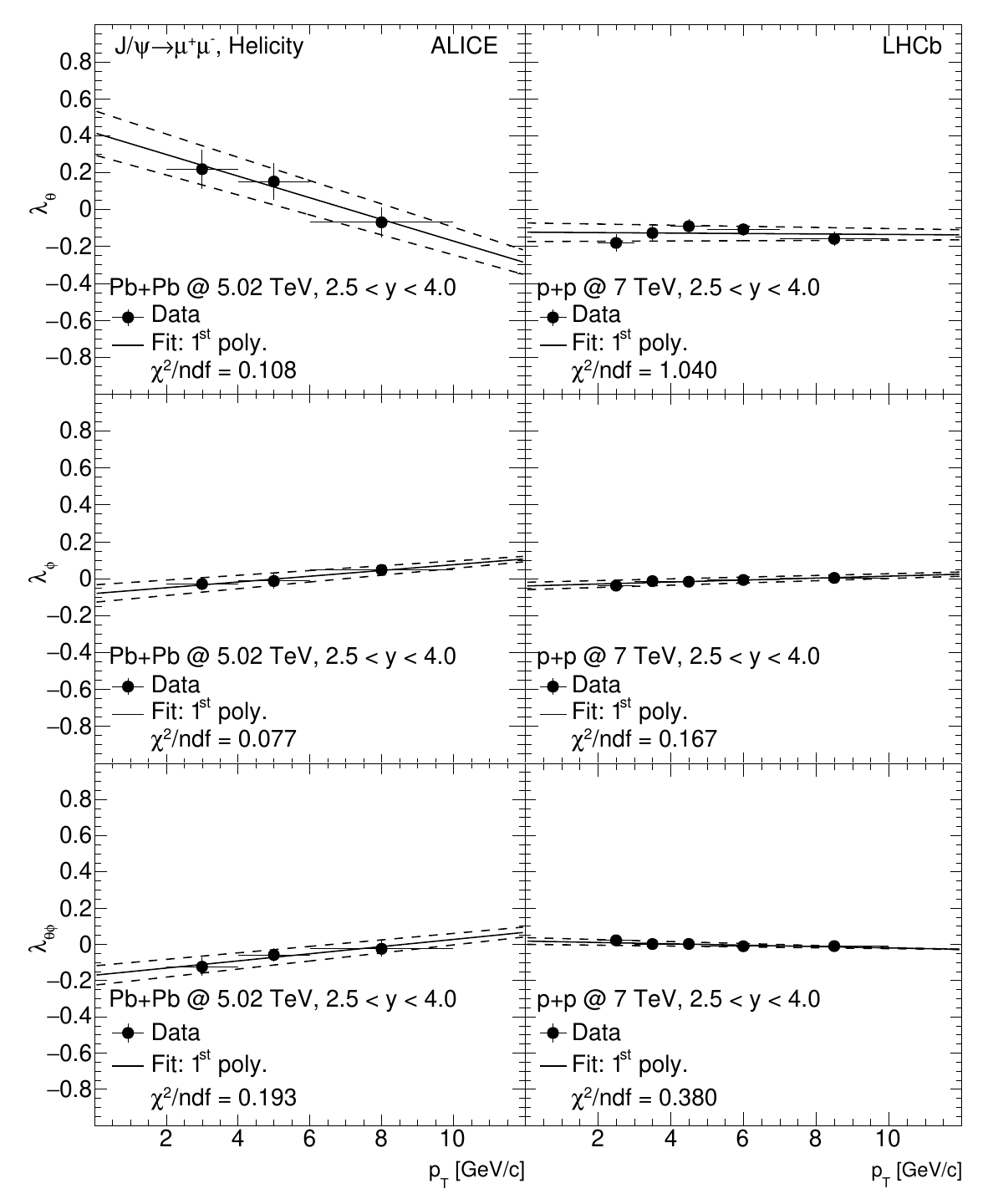}
    }
     \subfigure[]{
      \label{fig:cs_polar_lhc}
      \includegraphics[width=0.47\textwidth]{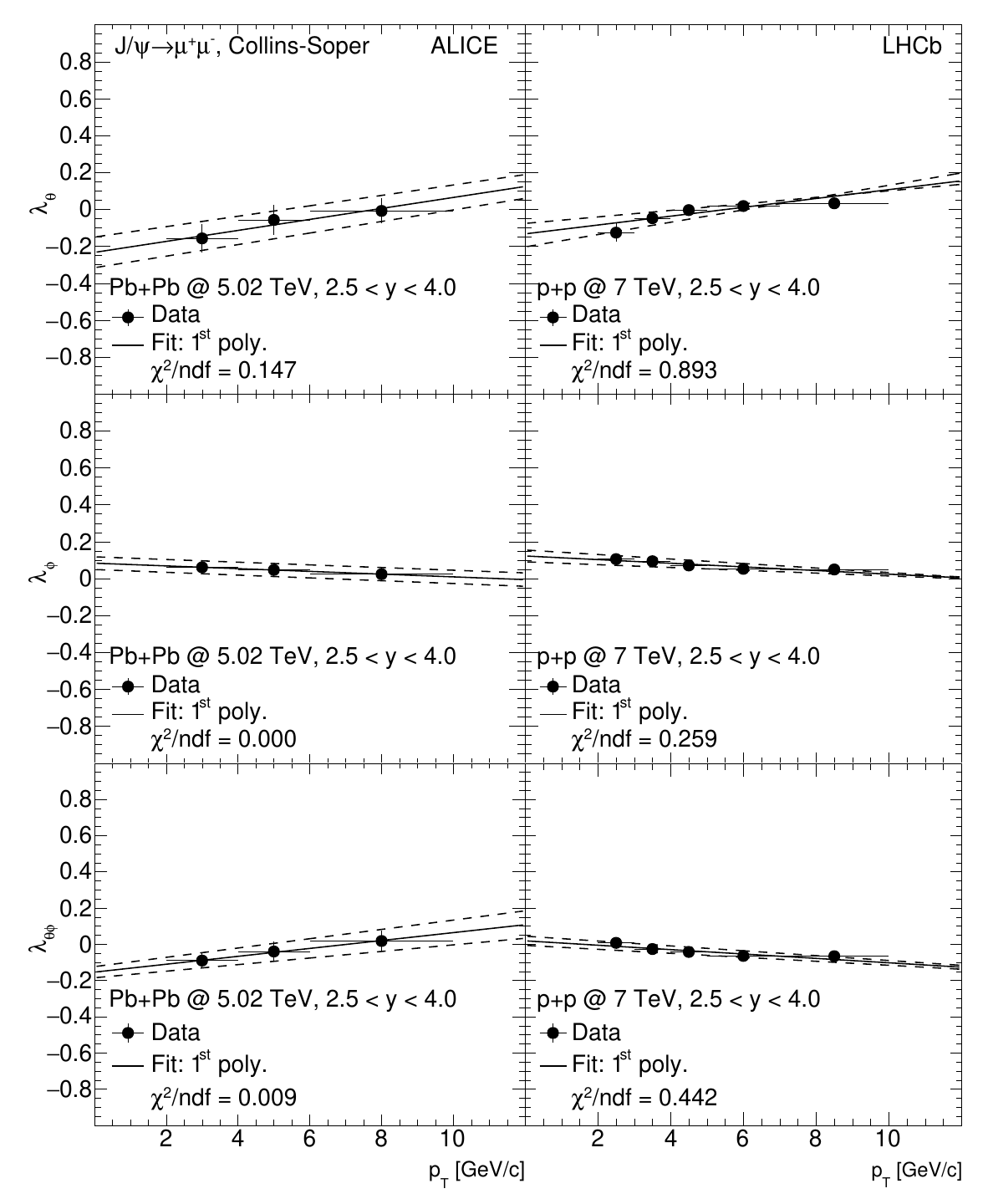}
    }  
  \end{center}
    \caption{ The polarization parameters $\lambda_{\theta}$, $\lambda_{\phi}$, and $\lambda_{\theta\phi}$ as a function of $p_T$ in \subref{fig:hx_polar_lhc} the HX and \subref{fig:cs_polar_lhc} CS frame measured from ALICE in Pb+Pb collisions at 5.02 TeV~\cite{alice_jpsi_pol_pbpb} and LHCb in $p$+$p$ collisions at 7 TeV~\cite{lhcb_jpsi_pol_pp}.
The rapidity range for these measurements is $2.5 < y < 4.0$. Black lines are a linear function fits to the data points and dashed lines are linear functions fit to the upper and lower bound of the date points. Note that the data points from LHCb are plotted only in the same $y$ region as ALICE.
\label{fig:fit_pol}}
\end{figure}

Figures~\ref{fig:hx_polar_lhc} and~\ref{fig:cs_polar_lhc} show the polarizations of $J/\psi$ measured by ALICE in Pb+Pb~\cite{alice_jpsi_pol_pbpb} and LHCb in $p$+$p$ collisions~\cite{lhcb_jpsi_pol_pp}.

\begin{figure}[!htbp]
  \begin{center}
      \subfigure[]{
     \label{fig:po_vs_unpo_hx}
     \includegraphics[width=0.45\textwidth]{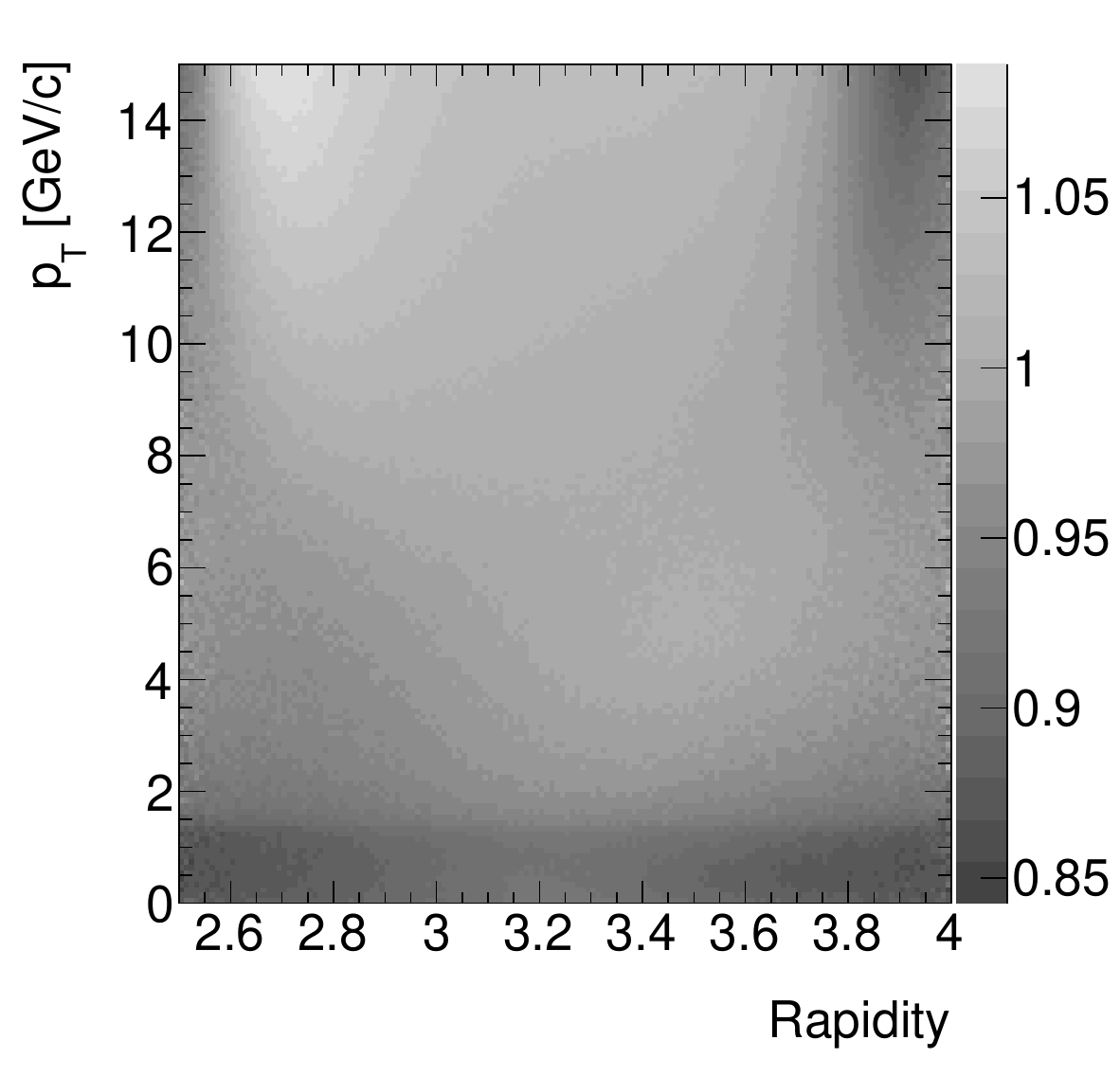}
    }
    \subfigure[]{
    \label{fig:po_vs_unpo_cs}
     \includegraphics[width=0.45\textwidth]{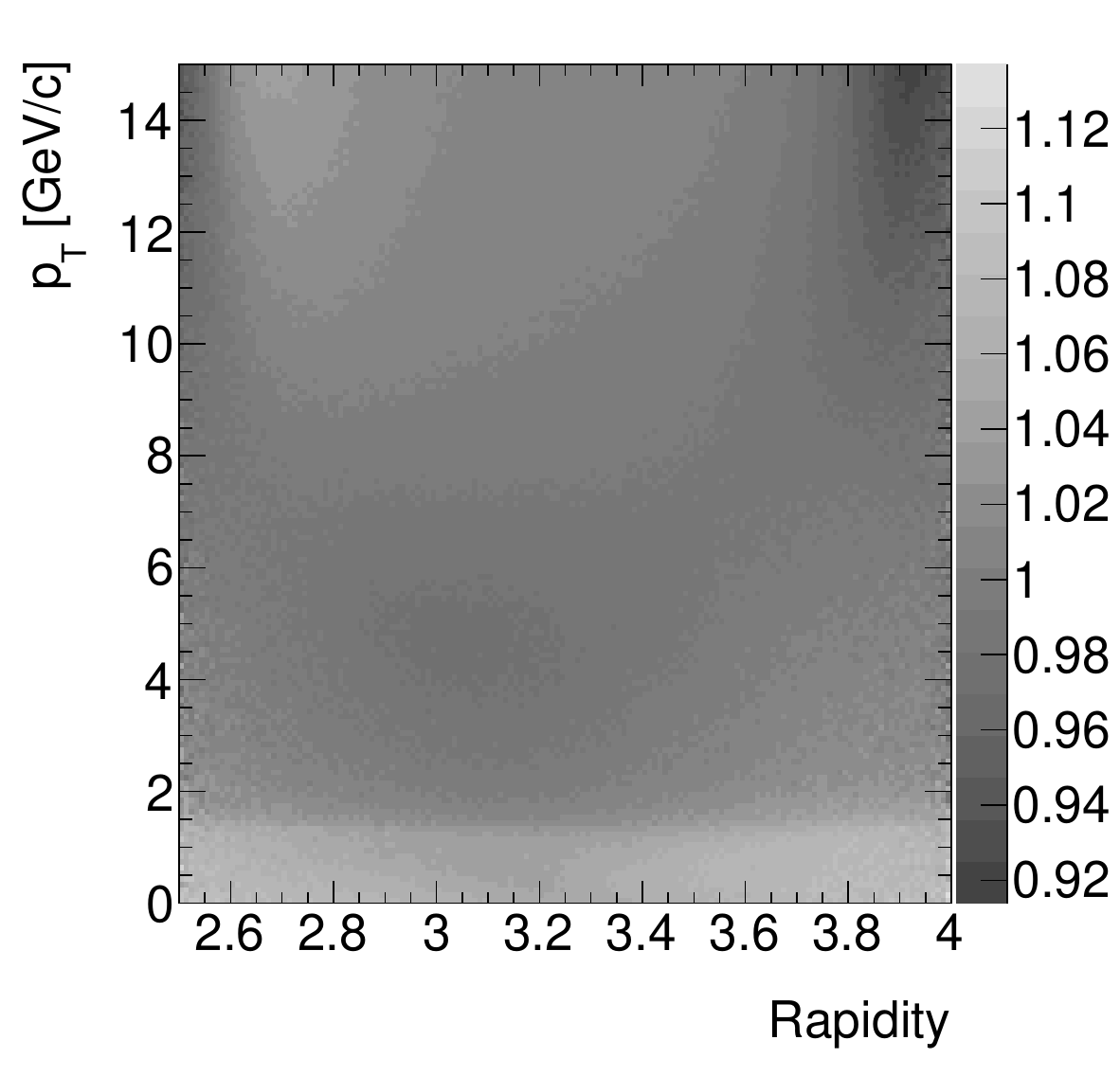} 
    } 
        \end{center}
    \caption{ The 2-dimensional ($J/\psi$ $p_T$ versus $y$) kinematic acceptance ratio of polarized $J/\psi$ with the parameters from the ALICE measurements to the unpolarized $J/\psi$ in \subref{fig:po_vs_unpo_hx} HX and \subref{fig:po_vs_unpo_cs} CS frame.
\label{fig:po_vs_unpo}}
\end{figure}

Figures~\ref{fig:po_vs_unpo_hx} and~\ref{fig:po_vs_unpo_cs} show the 2-dimensional kinematic acceptance ratio of polarized $J/\psi$ to unpolarized $J/\psi$ in the HX and CS frames. The kinematic acceptance varies significantly by $\sim$15\% ($\sim$12\%) in the low $p_T$ region and $\sim$8\% ($\sim$8\%) in the high $p_T$ region in the HX (CS) frame.

\begin{figure}[!htbp]
  \begin{center}
      \subfigure[]{
     \label{fig:jpsi_input_pt}
     \includegraphics[width=0.45\textwidth]{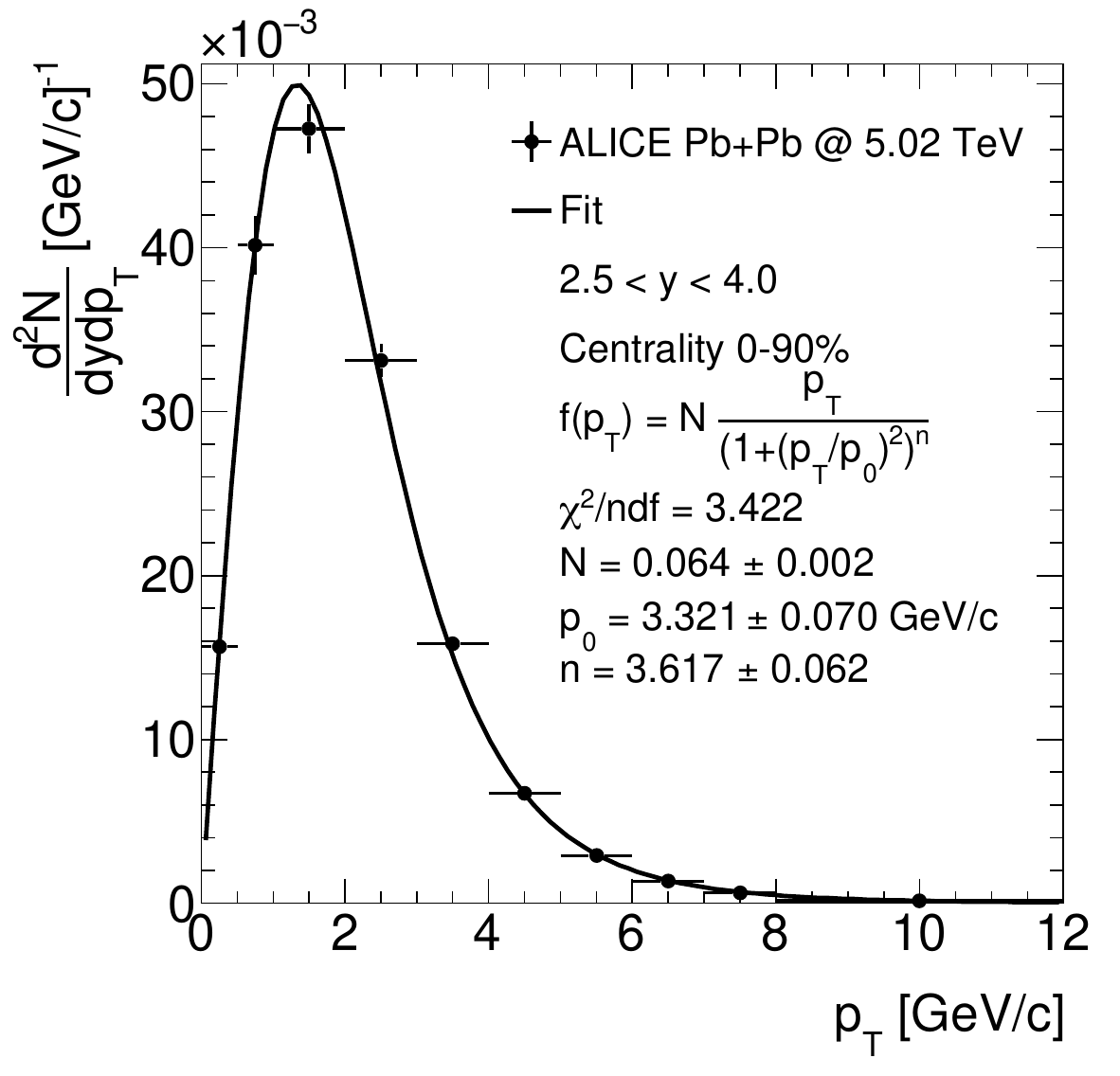}
    }
    \subfigure[]{
    \label{fig:correct_spectrum}
     \includegraphics[width=0.45\textwidth]{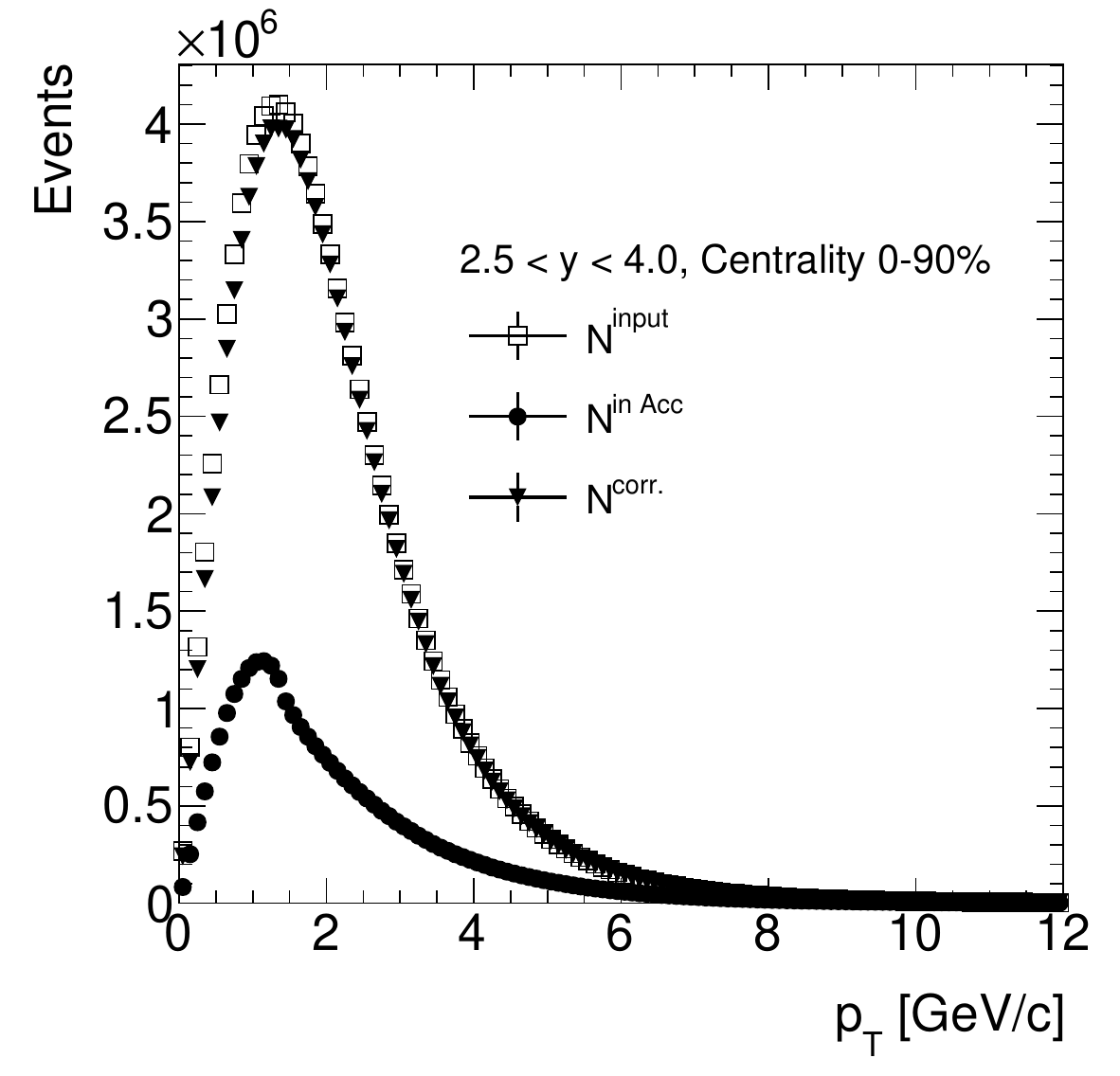} 
    } 
        \end{center}
    \caption{ \subref{fig:jpsi_input_pt} The $p_T$ spectrum of $J/\psi$ for 0-90\% centrality fitted by the function (black line) and details are described in the text.
\subref{fig:correct_spectrum} The $p_T$ spectra of the $J/\psi$ meson: the input spectrum ($N^{\rm input}$, open boxes), the spectrum with kinematic selections on $J/\psi$ and muons ($N^{\rm in\ Acc.}$, solid black circles), and the corrected spectrum using unpolarized acceptance ($N^{\rm corr.}$, solid triangles).
\label{fig:pt_alice}}
\end{figure}

Two key factors in Eq.~\ref{eq:raa_corr}, $N^{\rm input}$ and $N^{\rm corr.}$, are obtained from the MC events. In this procedure, the $y$ distribution in $A$+$A$ collisions is assumed to be identical to that in $p$+$p$ collisions. While experiments such as ALICE have observed a clear rapidity dependence of $R_{\rm AA}$ in Pb+Pb collisions~\cite{alice_raa_ydep}, this assumption serves as a necessary, controlled baseline to exclusively isolate the polarization effect without coupling it to rapidity modification factors. Figure~\ref{fig:pt_alice} shows the input $p_T$ spectrum of $J/\psi$ parameterized by $f(p_T) = N\frac{p_T}{(1+(p_T/p_0)^2)^n}$ alongside the corrected spectra. 

It must be explicitly noted that for certain datasets (such as the CMS Pb+Pb spectrum discussed later in Fig.~\ref{fig:jpsi_input_pt_cms}), this empirical fit yields large statistical uncertainties in its parameters due to the highly constrained functional form and limited data points at low $p_T$. However, this analytical formulation serves strictly as a purely empirical sampling weight for the Toy MC to replicate the transverse momentum shape. Because the kinematic acceptance effect is evaluated as a relative ratio ($N^{\rm input}/N^{\rm corr.}$), the absolute normalization $N$ cancels out entirely in Eq.~\ref{eq:raa_corr}. The mathematical instability of the fit parameters does not propagate into the final physical acceptance ratio.

\begin{figure}[!htbp]
      \begin{center}
      \includegraphics[width=0.9\textwidth]{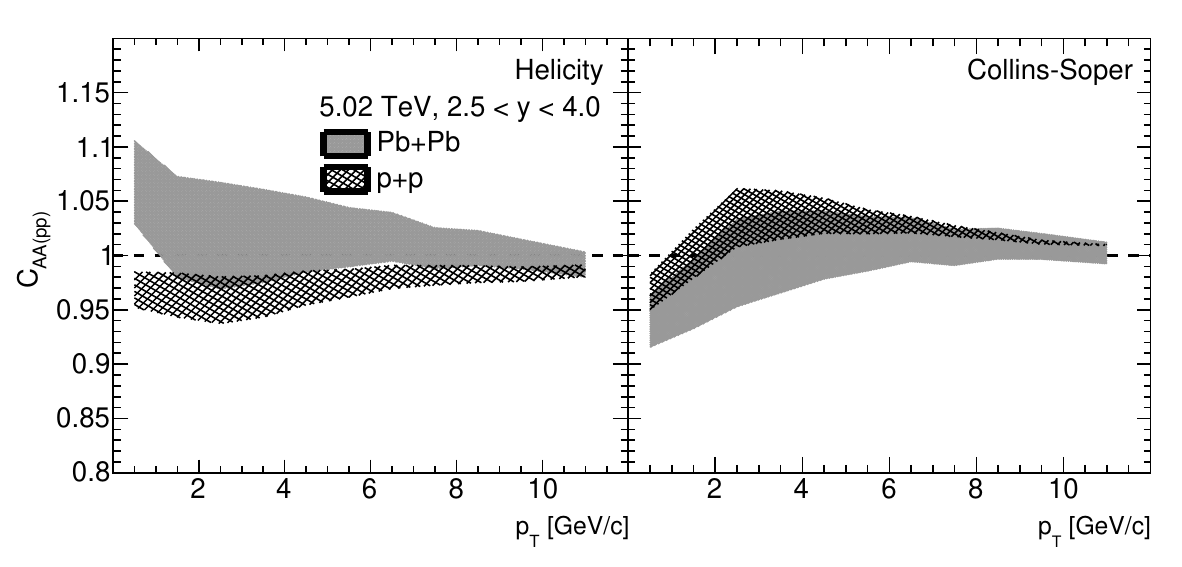}
      \end{center}
    \caption{ The $C_{\rm AA (pp)}$ as a function of $p_T$ in the HX (left) and CS (right) frame in the forward rapidity range ($2.5 < y < 4.0$) in Pb+Pb collisions at 5.02 TeV. The shaded area is due to the uncertainty of the polarization measurements.
\label{fig:ratio_forward_lhc}}
\end{figure}

Figure~\ref{fig:ratio_forward_lhc} shows the $C_{\rm AA (pp)}$ as a function of $p_T$. The shaded area reflects the systematic limit derived from the polarization measurements' uncertainties. 

\begin{figure}[htbp]
  \begin{center}
          \includegraphics[width=0.47\textwidth]{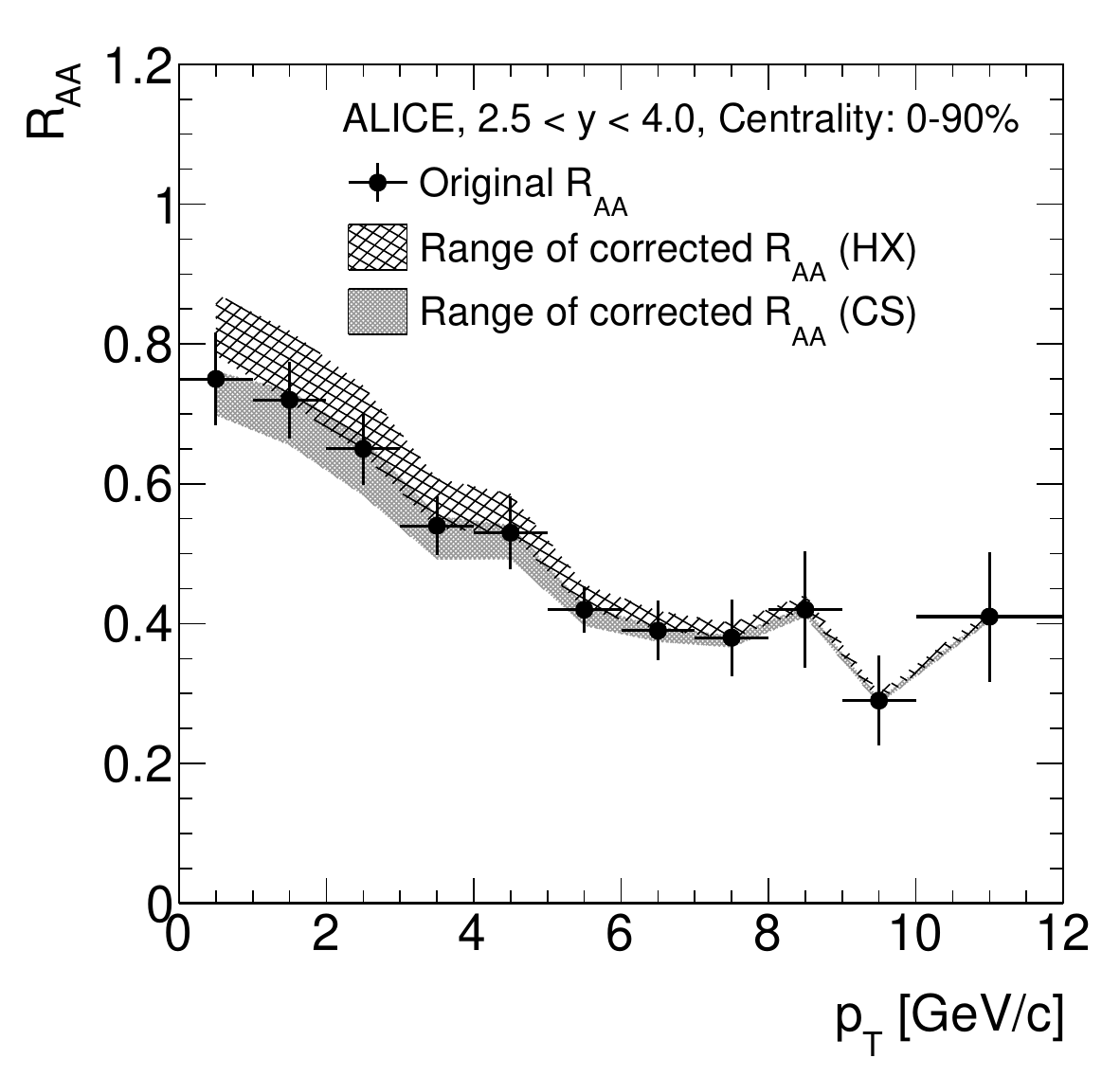}
      \end{center}
  \caption{ The $R_{\rm AA}$ measurements from ALICE (black points) and the corrected ones (shaded bands) in Pb+Pb collisions at 5.02 TeV in the HX and CS frame as a function of $p_T$.
\label{fig:final_results}}
\end{figure}

Finally, the original $R_{\rm AA}$ measurements from ALICE~\cite{alice_raa} and the modified sensitivity limits $R_{\rm AA}^{\rm corr.}$ are shown in Fig.~\ref{fig:final_results}. The result proves that the systematic deviation in the low $p_T$ region is not negligible (up to $\sim$16\%), highlighting a critical unquantified systematic uncertainty in existing interpretations.

\subsection{Central rapidity region}
\begin{figure*}[!htbp]
  \begin{center}
     \subfigure[]{
      \label{fig:fit_pol_star}
      \includegraphics[width=0.47\textwidth]{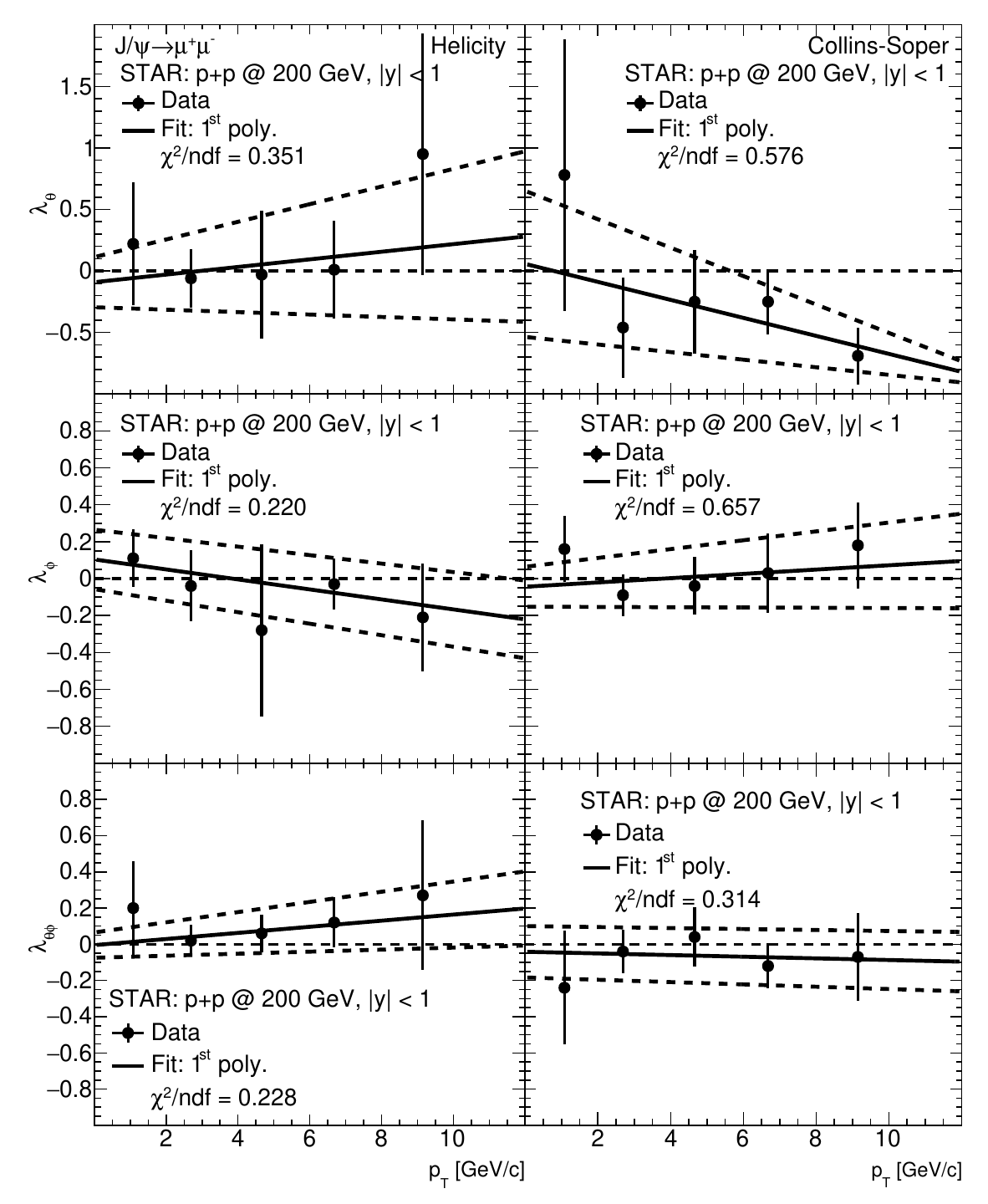}
    }
    \subfigure[]{
      \label{fig:fit_pol_cms}
      \includegraphics[width=0.47\textwidth]{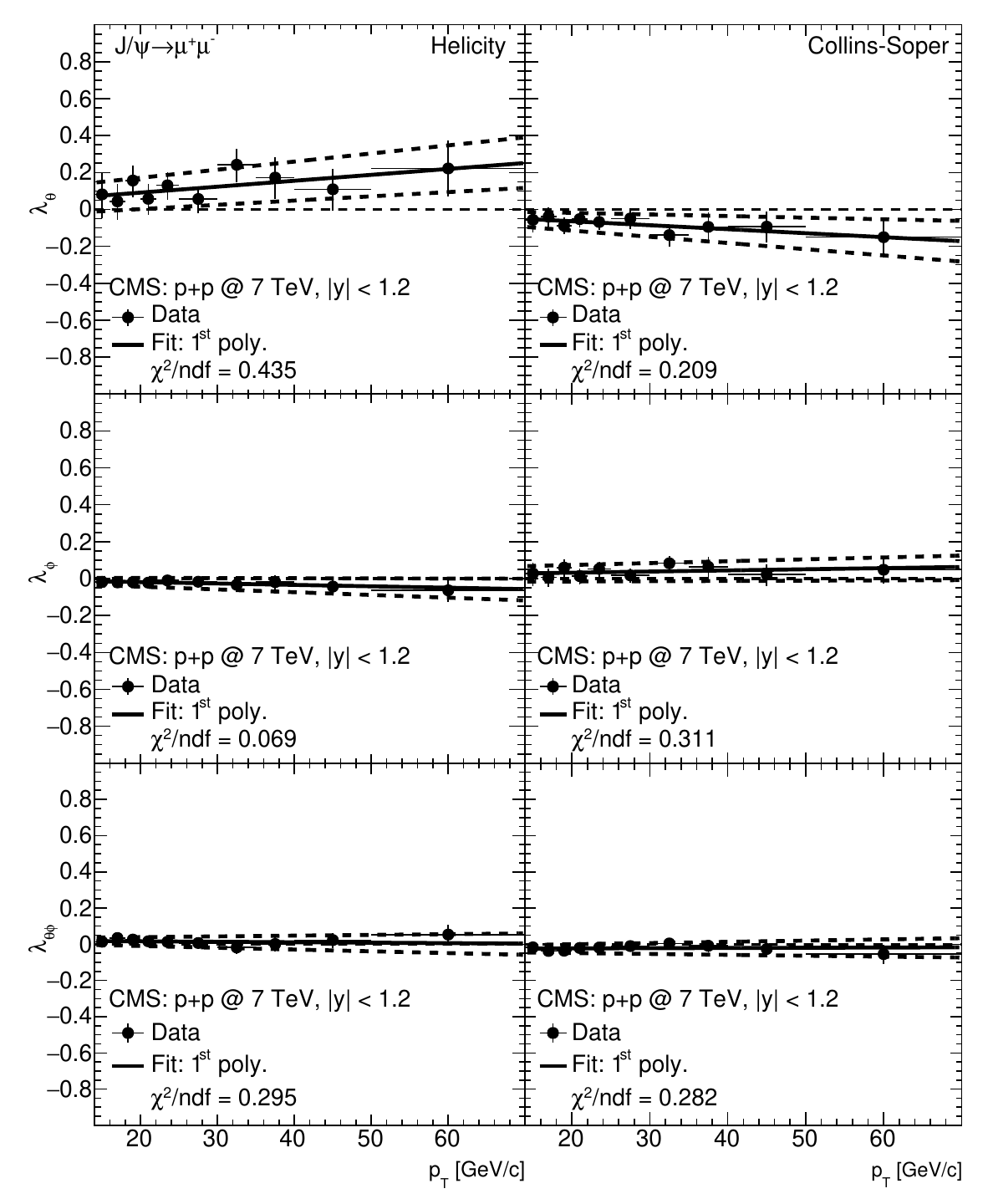}
    }  
  \end{center}
    \caption{ The polarization parameters $\lambda_{\theta}$, $\lambda_{\phi}$, and $\lambda_{\theta\phi}$ in $p$+$p$ collisions as a function of $p_T$ measured from the~\subref{fig:fit_pol_star} STAR at 200 GeV and~\subref{fig:fit_pol_cms} CMS at 7 TeV~\cite{star_pol, cms_pol}. The rapidity ranges for these measurements are $|y| < 1.0$ and $|y| < 1.2$ for STAR and CMS, respectively. Black lines are a linear function fits to the data points and dashed lines are linear functions fit to the upper and lower bound of the date points.
\label{fig:fit_pol_central}}
\end{figure*}

\begin{figure}[!htbp]
  \begin{center}
      \subfigure[]{
     \label{fig:jpsi_input_pt_star}
     \includegraphics[width=0.45\textwidth]{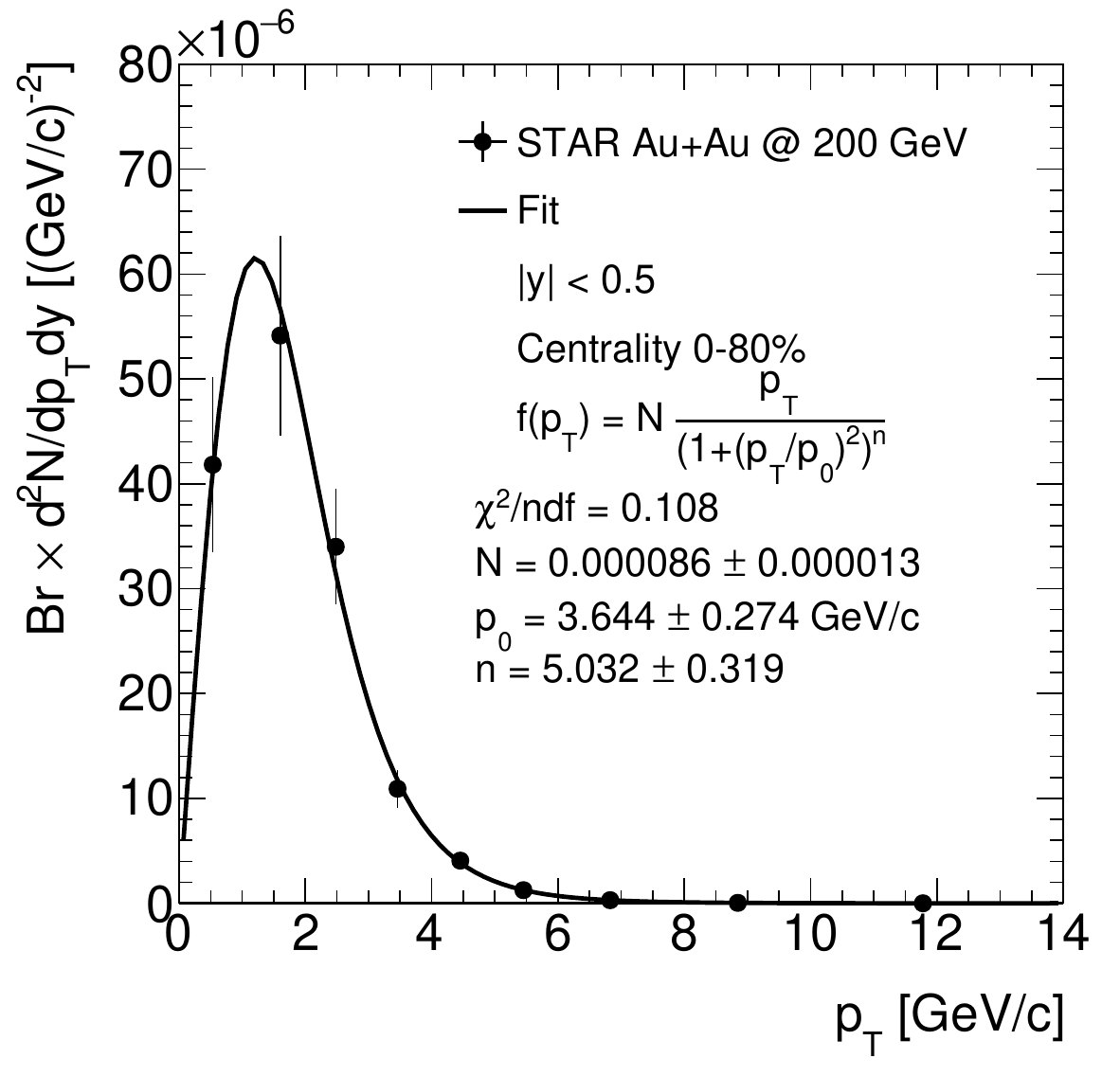}
    }
    \subfigure[]{
    \label{fig:jpsi_input_pt_cms}
     \includegraphics[width=0.45\textwidth]{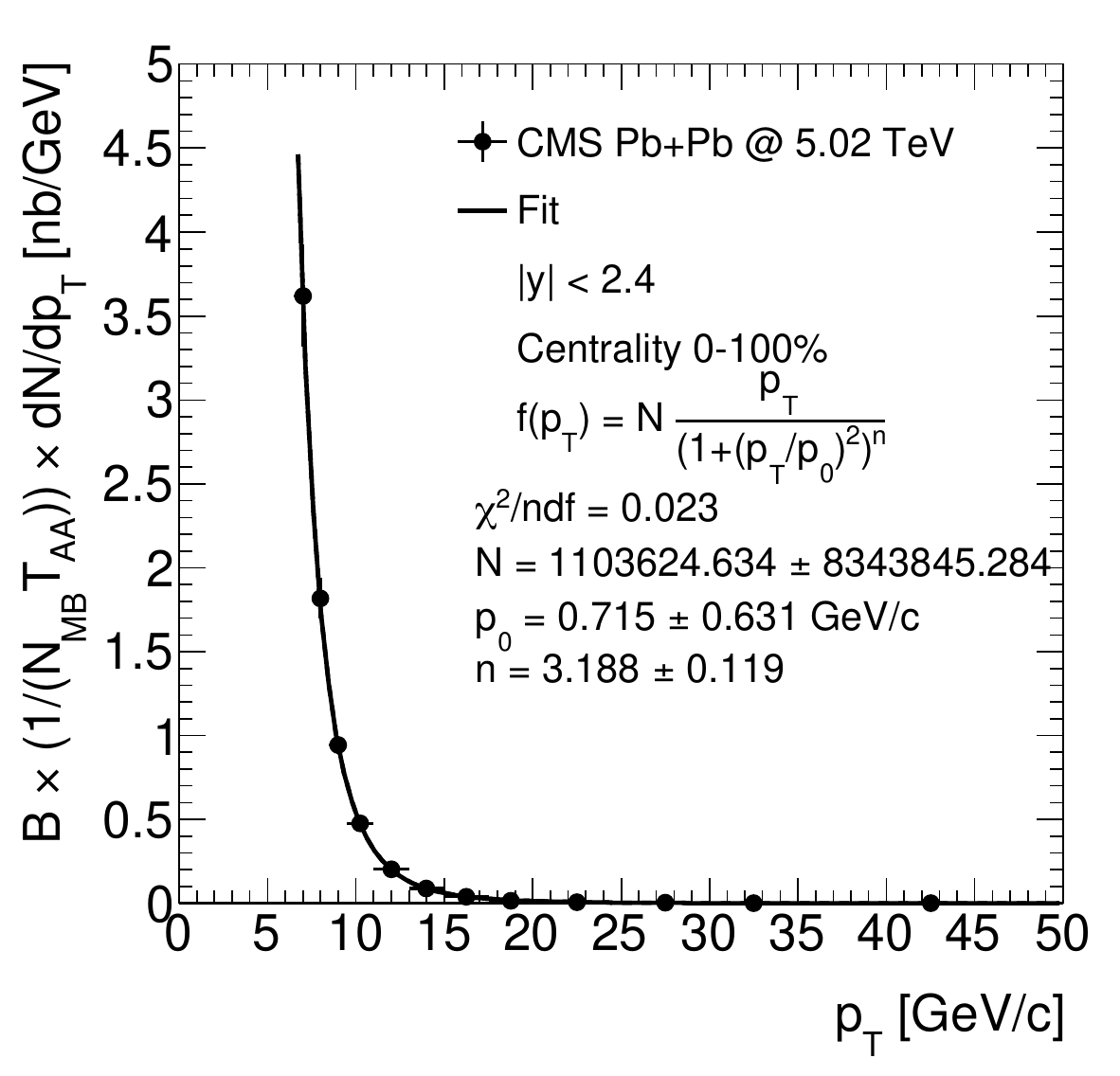} 
    } 
        \end{center}
    \caption{ \subref{fig:jpsi_input_pt_star} and \subref{fig:jpsi_input_pt_cms} are the $p_T$ spectra of $J/\psi$ from STAR in Au+Au collisions at 200 GeV and CMS in Pb+Pb collisions at 5.02 TeV.
\label{fig:pt_star_cms}}
\end{figure}

To establish the maximum systematic envelope of the QGP interpretation, we evaluated the central rapidity region. Polarization measurements from STAR~\cite{star_pol} and CMS~\cite{cms_pol} in $p$+$p$ collisions are utilized (Fig.~\ref{fig:fit_pol_central}). Since no direct polarization measurement of $J/\psi$ in heavy-ion collisions exists for this kinematic region, five extreme configurations are rigorously analyzed to cover the absolute boundaries of the polarization phase space: (1) unpolarized, (2) longitudinally polarized, (3) zero transversely polarized, (4) positively transversely polarized, and (5) negatively transversely polarized~\cite{atlas_jpsi, atlas_upsi, star_jpsi}. While it is acknowledged that such extreme, fully polarized states are physically unlikely to be fully realized in heavy-ion collisions, exploring these absolute boundaries is mathematically necessary to demonstrate the maximum possible phase space of the systematic uncertainty when experimental inputs are completely absent.

\begin{figure}[!htbp]
      \begin{center}
      \includegraphics[width=0.9\textwidth]{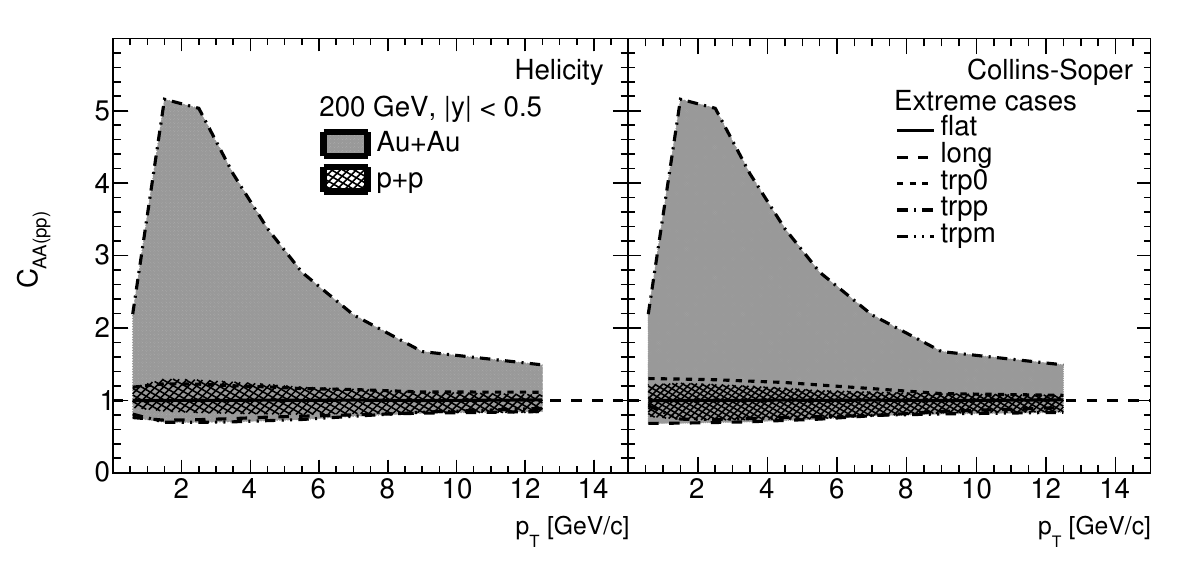}
      \end{center}
    \caption{ The $C_{\rm AA (pp)}$ as a function of $p_T$ in the HX (left) and CS (right) frame in the central rapidity range ($|y| < 1.0$) in Au+Au collisions at 200 GeV. The shaded area is due to the uncertainty of the polarization measurements. Different lines depict different polarization assumptions: ``flat'' is unpolarized, ``long'' is longitudinally polarized, ``trp0'' is zero transversely polarized, ``trpp'' is positively transversely polarized, and ``trpm'' is negatively transversely polarized.
\label{fig:ratio_central_star}}
\end{figure}

The systematic correction boundaries $C_{\rm AA (pp)}$ as a function of $p_T$ for Au+Au collisions at $\sqrt{s_{\rm NN}} = $ 200 GeV are shown in Fig.~\ref{fig:ratio_central_star}. Crucially, the extreme deviation limit in Au+Au collisions at this kinematic region is aggressively large, up to a factor of 6 in the low-$p_T$ region ($<$ 3 GeV). 

\begin{figure}[!htbp]
      \begin{center}
      \includegraphics[width=0.47\textwidth]{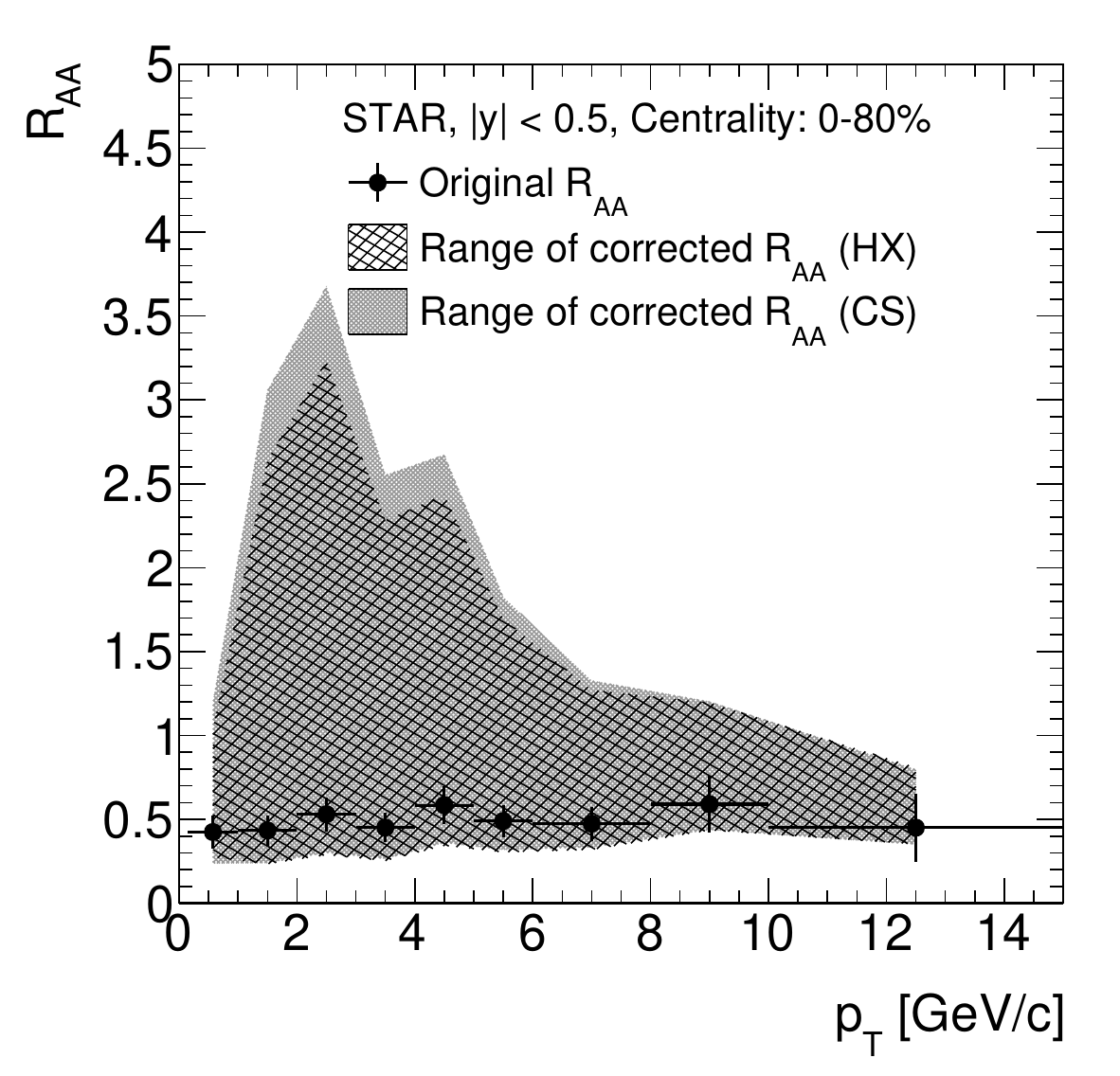}
      \end{center}
    \caption{ The $R_{\rm AA}$ measurements from STAR (black points) and the corrected ones (shaded bands) in Au+Au collisions at 200 GeV in the HX and CS frame as a function of $p_T$.
\label{fig:raa_star}}
\end{figure}

Figure~\ref{fig:raa_star} maps these systematic boundaries against the original $R_{\rm AA}$ measured by STAR~\cite{star_raa_orig}. This proves mathematically that without direct polarization measurements, the existing $R_{\rm AA}$ at low $p_T$ lacks the precision necessary to uniquely distinguish HNM from CNM effects.

\begin{figure}[!htbp]
      \begin{center}
      \includegraphics[width=0.9\textwidth]{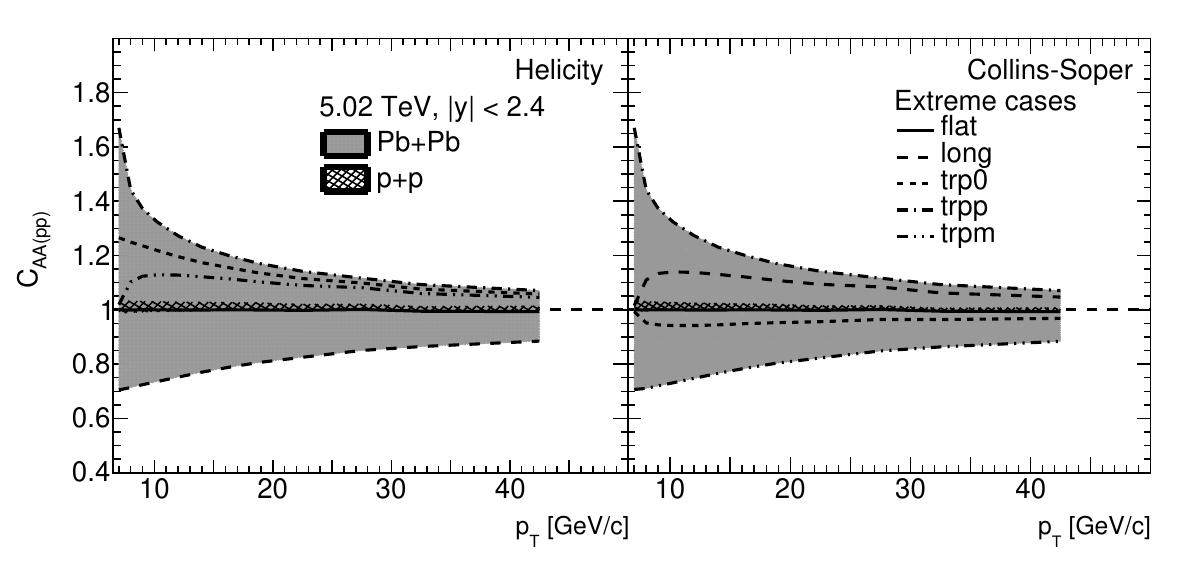}
      \end{center}
    \caption{ The $C_{\rm AA (pp)}$ as a function of $p_T$ in the HX (left) and CS (right) frame in the central rapidity range ($|y| < 1.2$) in Pb+Pb collisions at 5.02 TeV.
\label{fig:ratio_central_cms}}
\end{figure}

\begin{figure}[!htbp]
      \begin{center}
      \includegraphics[width=0.47\textwidth]{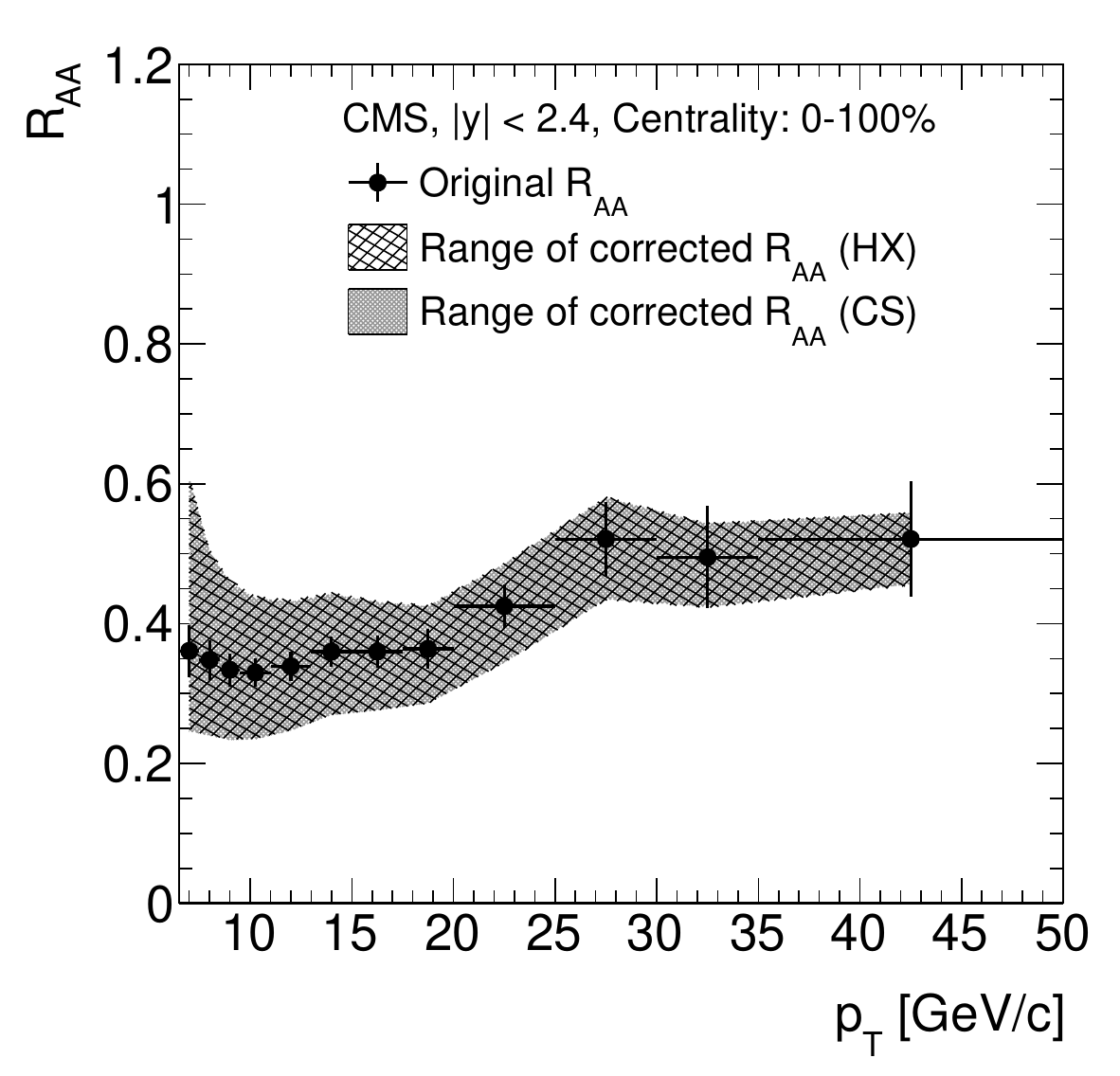}
      \end{center}
    \caption{ The $R_{\rm AA}$ measurements from CMS (black points) and the corrected ones (shaded bands) in Pb+Pb collisions at 5.02 TeV in the HX and CS frame as a function of $p_T$.
\label{fig:raa_cms}}
\end{figure}

Similar sensitivity limits are calculated for the LHC energy. The $C_{\rm AA (pp)}$ and the sensitivity envelope for CMS measurements~\cite{cms_raa_orig} are shown in Figs.~\ref{fig:ratio_central_cms} and~\ref{fig:raa_cms}. While the maximum deviation bounds are smaller at the LHC, they remain sizable ($\sim$$10\% - 70\%$), severely impacting the rigorous conclusion of any $R_{\rm AA}$ interpretation that assumes zero polarization.

\section{Conclusions}
To obtain an accurate interpretation to disentangle HNM and CNM contributions, all intrinsic dependencies of $R_{\rm AA}$ must be strictly quantified. The standard convention of calculating invariant yields under an unpolarized assumption introduces a notable systematic uncertainty.

This systematic sensitivity study demonstrates that based on recent ALICE and LHCb results, the unpolarized assumption induces a baseline systematic deviation of up to $\sim$16\% in the low $p_T$ regions where CNM contributes significantly. Furthermore, exploring the extreme physical boundaries of polarization in the central rapidity region reveals that the uncertainty envelope can expand up to a factor of 6 at RHIC energies and $\sim$70\% at LHC energies.

Conclusively, publishing and interpreting $R_{\rm AA}$ without considering the systematic uncertainty from unknown kinematic acceptance polarization parameters may lead to incomplete physical interpretations. Direct, precise measurements of quarkonium polarization in heavy-ion collisions are strictly required to validate the quantitative physical interpretations of heavy quarkonium interaction within the QGP.

\section*{Acknowledgments}
We thank the Institute of Physics, Academia Sinica, and National Cheng Kung University for their support. This work was supported in part by the Ministry of Science and Technology of Taiwan and Higher Education Sprout Project from Ministry of Education of Taiwan. We also thank Dr. Lijuan Ruan and Dr. Rongrong Ma from Brookhaven National Laboratory for useful suggestions.

\end{document}